\documentclass[traditabstract]{aa} 

\usepackage{txfonts} 
\usepackage{multirow}

\usepackage{graphicx}

\newcommand \ba{{ BATSE}}
\newcommand \fe{{\it Fermi}}
\newcommand \ep{$E_{\rm peak}$}
\newcommand \epo{$E^{\rm obs}_{\rm peak}$}
\newcommand \al{$\alpha$}
\newcommand \epop{$E^{\rm obs}_{\rm peak}$--$P$}
\newcommand \epof{$E^{\rm obs}_{\rm peak}$--$F$}
\newcommand \liso{$L_{\rm p,iso}$}
\newcommand \eiso{$E_{\rm iso}$}
\newcommand \gsim{ \lower .75ex \hbox{$\sim$} \llap{\raise .27ex \hbox{$>$}} } 
\newcommand \lsim{ \lower .75ex\hbox{$\sim$} \llap{\raise .27ex \hbox{$<$}} } 

\newcommand \ama{$E_{\rm peak}-E_{\rm iso}$}
\newcommand \yone{$E_{\rm peak}-L_{\rm iso}$}

\newcommand \F{$F$}
\newcommand \Pf{$P$}

\begin{document}
\title{\fe/GBM and \ba\ Gamma--Ray Bursts: comparison of the spectral properties}

\author{L. Nava \inst{1}\thanks{lara.nava@sissa.it}
 \and G. Ghirlanda \inst{2} \and G. Ghisellini \inst{2} \and A. Celotti \inst{1}}
\institute{SISSA, via Bonomea 265, I--34136 Trieste, Italy
\and Osservatorio Astronomico di Brera, via E. Bianchi 46, I--23807 Merate, Italy }
\date{Received .. ... .. / Accepted .. ... ..} 

\abstract{
The Gamma--ray Burst Monitor (GBM) on board \fe\ allows to study the spectra of Gamma Ray Bursts (GRBs) over an unprecedented 
wide energy range (8 keV -- 35 MeV). We compare the spectral properties of short and long GRBs detected by the GBM (up to March 2010) 
with those of GRBs detected by the \ba\ instrument on board the CGRO. GBM and \ba\ long bursts have similar distributions of fluence ($F$),   \epo\  
and peak flux ($P$) but GBM bursts have a slightly harder low--energy spectral index {\bf $\alpha$} with respect to \ba\ GRBs. 
GBM and \ba\ short bursts have similar distributions 
of fluence, $\alpha$ and peak flux, with GBM bursts having slightly larger \epo. We discuss these properties in light of the found correlations between \epo\ 
and the fluence and the peak flux. 
GBM bursts confirm that these correlations are not determined by instrumental selection effects.
Indeed, GBM bursts extend the \epof\ and \epop\ correlations both in fluence/peak flux and in peak 
energy.
No GBM long burst with  \epo\ exceeding a few MeV is found, despite the possibility of detecting it.
Similarly to what found with \ba, there are 3\% of GBM long bursts (and almost all short ones) that are outliers at more 
than 3$\sigma$ of the \ama\ correlation. Instead there is no outlier of the \yone\ correlation, for both long and short GBM bursts.
} 
\keywords{
Gamma-ray burst: general -- Radiation mechanisms: non-thermal }
\maketitle

\section{Introduction}

The \fe\ satellite, launched in June 2008, offers a great opportunity to 
characterize GRB spectra over a wide energy range thanks to its two high--energy instruments: 
the Large Area Telescope (LAT) and the Gamma--ray Burst Monitor (GBM -- Meegan et al. 2009). 
The GBM is composed by twelve NaI detectors (with good spectral resolution between 
$\sim$8 keV and $\sim$1 MeV) and two BGO detectors (operating between 200 keV and 40 MeV). 
Significant emission in the LAT energy range ($\sim$ 30 MeV -- 300 GeV) has been detected 
only in about 20 GRBs till now (December 2010), while the GBM triggered about 600 GRBs. 

A detailed spectral analysis of all GRBs detected by the \fe/GBM up to the end of March 2010 
(438 events) has been performed by Nava et al. (2010; N10 hereafter). 
These spectra were fitted with different models and for 316 events (272 long and 44 short) 
it was possible to constrain the peak energy of the $\nu F_\nu$ spectrum (\epo). 
For long bursts we found $\langle E^{\rm obs}_{\rm peak}\rangle\sim 160$ keV 
and an average low--energy power law index $\langle \alpha \rangle \sim$ --0.9. 
Short bursts are found to be harder, both in terms of \epo\ and $\alpha$:
$\langle E_{\rm peak}\rangle\sim$ 490 keV and $\langle \alpha \rangle \sim$ --0.5. 

N10 also analyzed the peak spectrum of GBM bursts, i.e. the spectrum corresponding 
to the peak flux of the light curve, accumulated on a timescale of 1.024 s and 0.064 
s for long and short events, respectively. 
The comparison with the time--integrated spectral properties shows that peak spectra, 
on average, have harder low--energy spectral indices but similar peak energies with 
respect to time--integrated spectra.

Before \fe, other instruments allowed to study the properties of the prompt emission spectra of GRBs. 
Due to its broad energy range ($\sim$25 keV -- $\sim$2 MeV), high sensitivity and detection rate, 
the \ba\ instrument onboard the {\it Compton Gamma Ray Observatory (CGRO)} 
satellite has been so far the instrument best suited to 
characterize the GRB prompt emission properties.
Thanks to its almost all--sky viewing, \ba\ detected more than 2700 GRBs in about 9 years.

The published spectral catalogs of \ba\ bursts comprise relatively small sub--samples of {\it bright} 
GRBs, selected on the basis of the burst fluence and/or peak flux 
(Preece et al. 2000; Kaneko et al. 2006 -- K06 hereafter). 
The analyzed samples allowed to study the spectral properties only of long bursts, 
given the small number of short bursts present in these samples (e.g. 17 short GRBs in the K06 sample). 
These studies revealed that the low--energy power law index distribution of long GRBs is 
centered around $\alpha\sim$ --1 and pointed out the inconsistency of the large majority 
of burst spectra with a synchrotron interpretation. 
The \epo\ distribution of long GRBs 
analyzed by K06 peaks around \epo$\sim$ 250 keV, with a relatively narrow dispersion. 
However, this refers to bright bursts, while fainter bursts have smaller 
\epo\ values, as shown by Nava et al. (2008, N08 thereafter).
They  performed the spectral analysis of a sample of \ba\ bursts 
selected by extending the limiting fluence of K06 (i.e.  $F=2\times 10^{-5}$ erg cm$^{-2}$) 
down to $F=10^{-6}$ erg cm$^{-2}$. 
They found that \epo\ correlates with the fluence $F$ and  the peak flux $P$. 
This sample of \ba\ faint bursts has a distribution of \epo\ values centered at 
$\sim$ 150 keV, i.e. a value smaller than 
the one found for the bright \ba\ bursts analyzed by K06, as a consequence
of the mentioned \epof\ correlation. 
This result confirms that the derived distribution of \epo\ is strongly 
affected by the adopted cuts in fluence (or peak flux).

N08 also found a correlation between \epo\ and the fluence/peak flux 
for short bursts.
This implies that when we compare the \epo\ 
distributions of short and long 
GRBs we must take into account the possible different fluence/peak flux selection criteria. 
A large sample of short BATSE bursts have been analyzed by  Ghirlanda et al. 2009 (G09 hereafter). 
They performed a detailed spectral analysis of  79 short bursts and compared their properties 
with those of 79 long BATSE bursts selected with the same limit on the peak flux. 
They found that the \epo\ distributions of the two classes are similar, while 
the low--energy power law indices are different:  
short bursts have $\langle \alpha\rangle \sim$ --0.4, harder than long events. 

A well known property of GRBs, related to their prompt emission, is the correlation between 
the rest frame peak energy \ep\ of long bursts with the bolometric isotropic energy \eiso\ 
emitted during the prompt (Amati et al. 2002) or with the bolometric isotropic luminosity 
\liso\ estimated at the peak of the light curve (Yonetoku et al. 2004). 
Such correlations represent  an intriguing clue on the dominant emission mechanism of the prompt phase. 
Furthermore, if corrected for the jet opening angle, their dispersion reduces considerably 
(Ghirlanda et al. 2004a) and allows to use GRBs as standard candles (Ghirlanda et al. 2004b). 

The correlations in the observer frames (\epof\ and \epop) may be just the consequence of 
the rest frame (\ama\ and \yone\ respectively) correlations mentioned above. 
Alternatively, it has been claimed that the rest frame correlations are the 
result of instrumental selection effects (Band \& Preece 2005; Nakar \& Piran 2005). 
Ghirlanda et al. (2008, hereafter G08) and N08 examined the instrumental selection 
effects which may affect the observer frame correlations \epof\ and \epop. 
They found that, although instrumental biases do affect the burst sample 
properties, they are not responsible for the correlations found in the observational planes.

Moreover, Ghirlanda et al. (2010a) recently showed that the correlation \ep--\eiso\  
and \ep--\liso\ holds for the time--resolved quantities within individual long GBM bursts 
(see also Firmani et al. 2009 for {\it Swift} bursts and Krimm et al. 2009 for 
{\it Swift--Suzaku} GRBs) 
and that this ``time--resolved" correlation is similar to that defined by the time--integrated properties of different GRBs. 
Similar results were found for short GBM bursts: there is a significant correlation 
between the observer frame peak energy \epo\ and the peak flux within individual 
short GRBs and this correlation has a slope similar to that of the rest frame 
\yone\ correlation (Ghirlanda et al. 2010b). 
These results confirm that the ``Amati" and ``Yonetoku" correlations have a physical origin, 
instead of being the result of instrumental selection biases as claimed, and that the 
trends (\epof\ and \epop) seen in the ``observational planes" are just their outcome.

Through the spectral catalog of GBM bursts of N10, we can study the distribution of 
GBM bursts in the observational planes \epof\ and \epop\ and test if the correlations 
found by BATSE bursts (G08, N08 and G09) still hold.
In the observational planes we can also study, 
for the first time, the possible instrumental biases of GBM for the bursts 
analyzed in N10, and compare the spectral properties of long and short GRBs detected by \ba\ and by the GBM. 
Finally, we can also compute the fraction of GBM short and long 
GRBs which are outliers, for any assigned redshift, of the rest frame \ama\ and \yone\ correlations. 
These are the main aims of the present paper. 

In \S 2 we present the samples of \ba\ and GBM bursts (both long and short) used for our comparison.
We compute (\S 3) the relevant instrumental selection effects introduced  
by the GBM on the observational \epop\ and \epof\ planes considering separately short and long GRBs.
The comparison between BATSE and GBM results is also presented in terms of 
spectral parameters distributions in \S 4. We discuss our results and draw our conclusions in \S 5.

\section{Samples}

\subsection{Long bursts}

\begin{figure} 
\vskip -0.5 cm
\hskip -1.1 cm
\includegraphics[scale=0.55]{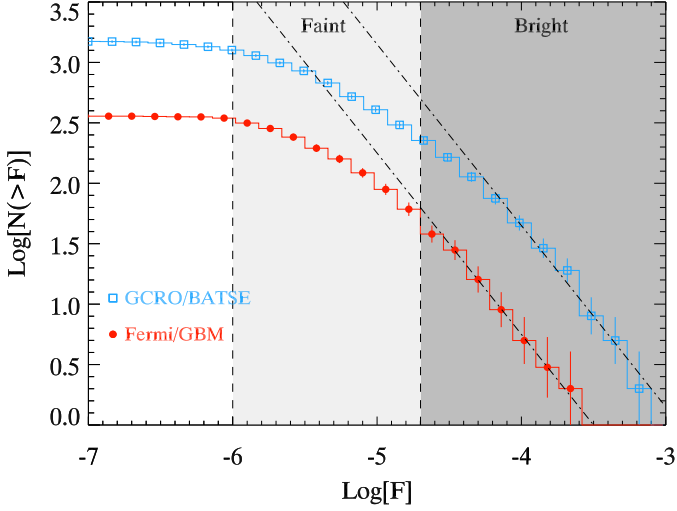}
\vskip -0.2 cm
\caption{{\it Long GRBs.} Log$N$--Log$F$ for long \ba\ bursts (empty squares) 
and long GBM bursts (filled circles). 
In both cases the fluence $F$ is integrated between 20 keV and 2000 keV. 
GBM data are taken form N10, while \ba\ data are from the online catalog and 
include all the \ba\ bursts for which the fluence has been estimated. For reference two power laws with slope -3/2 are shown (dot--dashed lines).}
\label{fluence} 
\end{figure}

{\it BATSE ---}
Fig. \ref{fluence} shows the Log$N$--Log$F$ of long GRBs detected by \ba\ (open squares), where
$N$ is the number of objects with fluence larger than $F$.
Since we compare this Log$N$--Log$F$ with the one of GBM bursts (filled squares in Fig. \ref{fluence}),
we compute $F$ in the energy range between 20 keV and 2000 keV, which is 
common to both instruments.
\ba\ fluences are taken from the online $CGRO$/BATSE Gamma--ray Bursts 
Catalog\footnote{http://heasarc.gsfc.nasa.gov/W3Browse/cgro/batsegrb.html}. 

We have then considered the spectral catalog of K06 
which contains all BATSE bursts with peak flux  
$P~(50-300 \rm~keV)>10~photons~cm^{-2}~s^{-1}$ {\it or} fluence 
$F~(20-2000\rm~keV)>2\times10^{-5} \rm~ergs~cm^{-2}$.  
From the 350 events of this sample we extract the long ones, i.e. with observed 
duration $T_{90}>2$ s. 
Among these, 280 GRBs have a well determined \epo\ (i.e. their spectra are best fitted by a curved model 
with $\alpha<-2$ or $\beta>-2$, i.e. showing a peak in a $\nu F_{\nu}$ representation). 
For 104 events the spectrum is best fitted by a BAND model (Band et al. 1993), for 65 by a 
COMP model (a power law with a high--energy exponential cutoff) and for 111 by a 
smoothly broken power law model (SBPL).

Although the K06 analysis enlarges the previous spectral catalog of \ba\ bursts (Preece et al. 2000) 
it still selects only the brightest \ba\ bursts (i.e. corresponding to only the 13\% of the 
entire population of bursts detected by \ba). 
N08 selected a sample of 100 \ba\ bursts with a fluence fainter than the threshold adopted by K06. 
The N08 events, in fact, have a fluence in the range $10^{-6} <$ $F$ $<2\times10^{-5}\rm~erg~cm^{-2}$. 
Moreover, although these are only 100 bursts, they are representative of the large population 
of $\sim$1000 bursts in this fluence range since they were randomly extracted 
following the Log$N$--Log$F$ distribution of \ba\ GRBs in this fluence range. 
Of these 100 representative bursts, 44 are best fitted by the COMP model, 44 by the BAND model, 
and 12 by a power law (PL) function. 
Therefore the N08 sample contains 88 GRBs with a spectrum fitted by a curved model, 
for which \epo\ was well determined. Among the bursts analyzed by K06 with peak flux $P$ larger than 10 ph cm$^{-2}$ s$^{-1}$,
there are GRBs with fluences smaller than $F = 2\times10^{-5}$ erg cm$^{-2}$,
that overlap with the ones studied by N08.
We exclude these bursts from the present discussion, in order
to have well defined complete samples at two limiting fluences,
that we will call, in the rest of the paper, 
the ``bright" \ba\ bursts (the bursts in K06 with $F > 2\times10^{-5}$ erg cm$^{-2}$)
and the ``faint" \ba\ bursts (the bursts studied in N08 with
$10^{-6} <F< 2\times10^{-5}$ erg cm$^{-2}$).
The grey shaded regions in Fig. \ref{fluence} correspond to this subdivision.

{\it GBM ---}
In N10 we have analyzed the spectra of all the GRBs detected by the GBM up to 
March 2010 (438 GRBs). 
No fluence or peak flux selection has been adopted. 
Fig. \ref{fluence} shows that the
shape of the Log$N$--Log$F$ for the two instruments is very similar. 
To compare the \epo\ and $\alpha$ 
distribution of GBM bursts with those of \ba\ bursts, we select from the N10 catalog 
two subsamples with the same fluence criterion adopted by K06 and N08 for \ba\ bursts 
and we call them the \emph{bright GBM sample} and the \emph{faint GBM sample}. 

In six GBM bursts N10 could not analyze the spectrum, due to lack of data. 
For the remaining 432 bursts we performed the spectral analysis using a power law model (PL -- 110 spectra), 
a COMP (232  spectra) and a Band model (90 spectra) and evaluating for each burst the spectral parameters 
of the best fit model. 359 events belong to the long burst class. We also estimated their peak flux on time bin 
of 1.024 seconds. In this work we will use this sample of GBM long bursts for comparison with \ba\ long GRBs.

\subsection{Short bursts}

{\it BATSE ---}
The most comprehensive sample of short \ba\ GRBs with well defined spectral parameters 
is composed by the 79 events analyzed by G09, selected for having $P >3~\rm photons~cm^{-2}~s^{-1}$. 
In 71 cases the spectra have a well determined \epo.

{\it GBM ---}
For the GBM instrument we use the spectral parameters of the 44 short 
GRBs present among the 438 bursts analyzed by N10 with a well determined \epo. 
Their peak flux is estimated on a time bin of 0.064 seconds.

\section{\epo--Fluence and \epo--Peak Flux planes: comparison between BATSE and GBM bursts}
\label{piani osservativi}

A correlation between the total fluence and \epo\ 
was first found by Lloyd, Petrosian \& Mallozzi (2000) for a sample of \ba\  
bursts without measured redshifts.  
This finding was recently confirmed 
by Sakamoto et al. (2008) using a sample of bursts detected by Swift, BATSE and Hete--II. 
In particular, they noted that X--Ray Flashes and X--Ray Rich bursts 
satisfy and extend this correlation to lower fluences. 

The distribution of GRBs with and without measured redshift in the planes 
\epo--\F\ and \epo--\Pf\ has been investigated by G08 and N08.
N08 considered all events with published spectral 
information detected by different instruments ({\it Swift}, BATSE, {\it Hete--II}, 
Konus/Wind and {\it Beppo}SAX) together with the 100 faint BATSE bursts analyzed in that paper. 
In both planes long bursts define a correlation, with fainter bursts having lower \epo.

In order to examine the distribution of GBM bursts in the observational planes 
\epo--\F\ and \epo--\Pf\ and compare it with the BATSE bursts we have first to
estimate the possible instrumental biases induced by the detector (see G08). 
One instrumental bias is the capability of an instrument 
to be triggered by a burst, i.e. the ``trigger threshold" (TT) (first computed by 
Band 2003 for different detectors). 
The second bias concerns the minimum number of photons required to analyze the spectrum and 
constrain the spectral parameters. 
This is called ``spectral threshold" (ST) in G08 and N08. 
The TT translates into a minimum peak flux, which depends on the burst spectrum and 
in particular on its \epo\ and can be described as a curve in the \epo--\Pf\ plane.
The second requirement (ST) results into a minimum fluence which depends on  
\epo\ and also on the burst duration. 
For this reason the ST is represented as a region (i.e. not a line) in the \epo--\F\ plane.

These curves (TT and ST) divide the observational planes \epof\ and \epop\ into two regions. Bursts with 
peak energy and peak flux which place them on the left of the TT curve cannot be triggered by the corresponding
instrument. Similarly, bursts with peak energy and fluence which place them on the left side of the ST curves do not
have enough photons to allow a reliable spectral analysis (see G08 for more details).

\begin{figure*} 
\hskip 0.3 truecm
\includegraphics[scale=0.57]{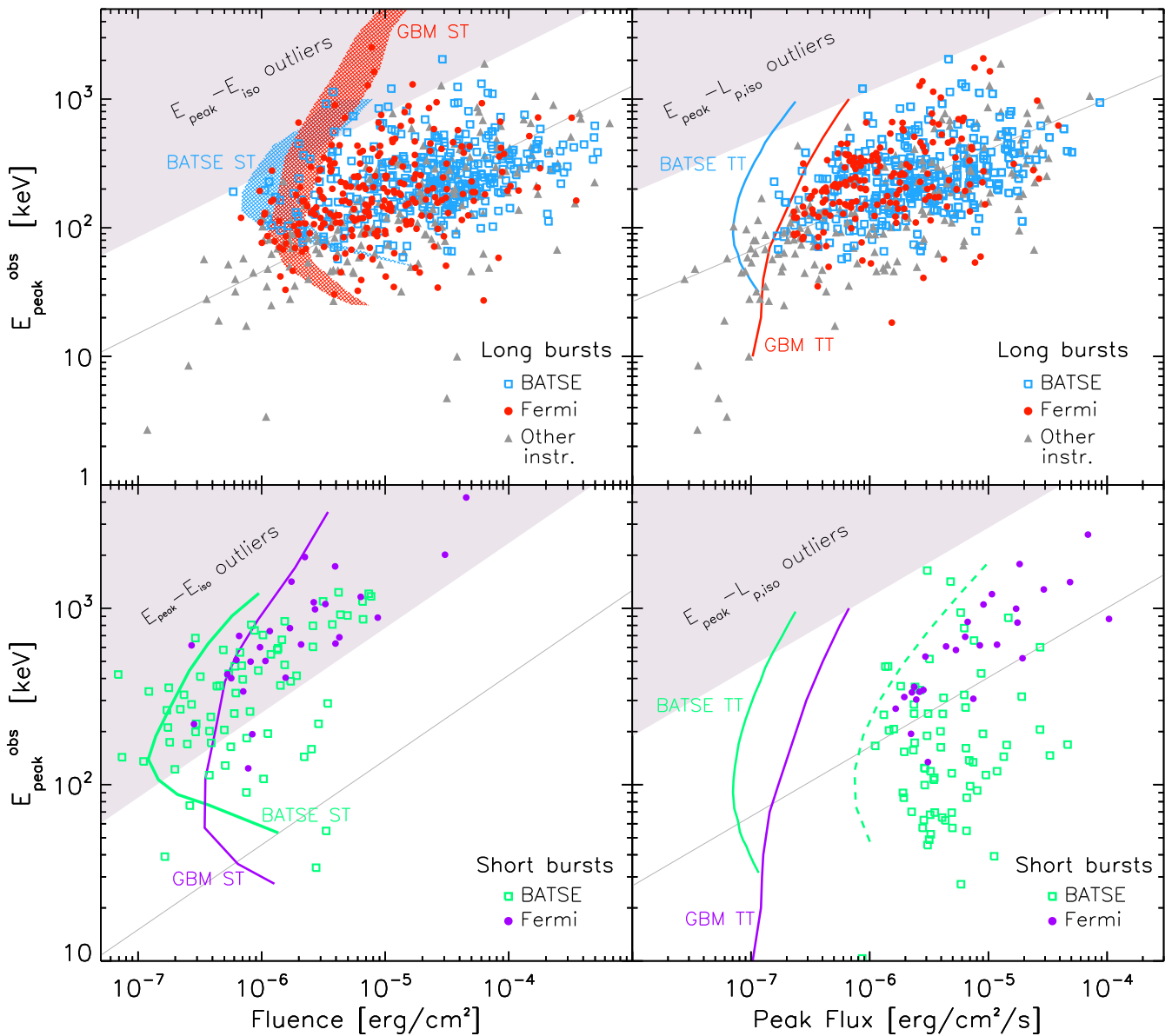}
\caption{
\epo--Fluence and \epo--Peak Flux planes for long (upper panels) 
and short (bottom panels) bursts. 
Empty squares represent BATSE bursts, filled circles GBM bursts and filled 
triangles indicate events detected by other instruments (from N08). 
In all panels the instrumental limits for BATSE and GMB are reported: 
shaded curved regions in the upper left panel show the ST, estimated assuming 
a burst duration of 5 and 20 s; 
solid curves in the bottom left panel represents the ST for short bursts. 
Solid curves in the right panels define the TT, identical for short and long events. 
Thresholds for BATSE are taken from G08 while those for the GBM instrument 
are derived in this work (see \S 3.1).
The dashed curve in the bottom right panel represents the selection 
criterion applied by G09 for their sample of short bursts, i.e. $P>$ 3 phot cm$^{-2}$ s$^{-1}$.
The shaded regions in the upper left corners of all the planes are the region identifying 
the outliers at more than 3$\sigma$ of the \ama\ (left panels) and \yone\ (right panels) 
correlations for any given redshift. 
GRBs, without measured redshift, which fall in these 
regions are outliers of the corresponding rest--frame correlations (\ama\ and \yone\ for 
the left and right panels respectively) for any assigned redshift. 
It means that there is no redshift which make them consistent with these 
correlations (considering their 3$\sigma$ scatter).
} 
\label{piani correle} 
\end{figure*} 

\subsection{Estimate of GBM instrumental selection effects}

Following G08, the TT curves are obtained adapting the results of Band 
(2006) and are shown in the right (top and bottom) panels of Fig. \ref{piani correle}. 
The TT curves are the same for long and short bursts as the trigger 
threshold depends only on the peak flux. 

The ST curves have been calculated from numerical simulations, as described in  G08. 
To perform these simulations, the typical background spectrum and the 
detector response function are required. For the bursts detected by the GBM both of them depend on 
several factors (e.g. the satellite attitude when a burst occurs), and thus 
there is no universal background and response matrix which can be adopted. 
To overcome this problem we use the real 
backgrounds and responses of several GRBs detected by the GBM and 
average the results of the simulations to build average ST curves 
for the population of long and short bursts, respectively.

Our simulation performs a joint spectral analysis of spectra simulated for two  
NaI detectors and one BGO detector. For long bursts, simulations were performed 
using the detector response files and the background spectra of the long GBM bursts 
published in G10. We considered, as done in G08, two representative values of the duration i.e. 
$T_{90} =$ 5 and 20 s (corresponding in Fig. \ref{piani correle} to the curves 
delimiting the red shaded region on the left and right side, respectively).
For short bursts we estimated the ST curves adopting the response files and the background spectra of the 
short bursts of the GBM sample of N10 assuming a typical duration of the simulated spectra of 0.7 s 
(red curve in the bottom left panel of Fig. \ref{piani correle}). 
This value, also adopted by G08 for \ba\ bursts, corresponds to the  typical duration of short 
GRBs observed by the GBM. 

For the TT and ST curves of the BATSE instrument we simply report those obtained in  
G08 (for long bursts) and in G09 (for short bursts).

Fig. \ref{piani correle} shows the distribution of GBM bursts in the 
\epof\ and \epop\ planes (right and left panels) for long and short GRBs (upper and bottom panels).

\subsubsection{Long bursts}

\emph{\epo\ vs Fluence ---} 
As can be seen in the top left panel of Fig. \ref{piani correle}, the distribution of long
GBM bursts (filled circles) extends down to the lower end of the distribution of \ba\ bursts (empty squares).

The presence of GBM bursts with low \epo\ (between $\sim$10 and $\sim$50 keV), not present in the 
BATSE sample, is clearly due to the wider energy range of the GBM instrument, 
sensitive down to $\sim$8 keV (see the ST curves). 
In this region, GBM bursts are consistent with bursts detected by other instruments (filled triangles). 
We also note that GBM bursts define a correlation which mostly overlaps with that 
defined by BATSE bursts and extends to the lower--left part of the  \epo--\F\ plane.

Despite the GBM assures good coverage up to $\sim$30 MeV (vs $\sim$ 1 MeV of BATSE),
there are only a few long GRBs with \epo\ exceeding few MeV, similarly to what found for 
the population of \ba\ bursts. 
Note that high \epo\ also means GRBs with high fluences, which are rarer than GRBs with low fluences.

Note also that while BATSE bursts appear concentrated at the high end of the \epof\ correlation, 
the sample is composed by \emph{all} bursts with 
$F>2\times10^{-5}$ erg cm$^{-2}$ (the K06 sample) and only \emph{one hundred} of fainter 
bursts (the N08 sample) representative of $\sim1000$ objects with 
$10^{-6} < F < 2\times10^{-5}$ erg cm$^{-2}$. 
Therefore the real density of \ba\ bursts in the latter fluence range is much larger 
than what represented in Fig. \ref{piani correle} (see e.g. Fig. 8 in N08) so that 
the slight shift between \ba\ and GBM population density in the upper panels is only 
an apparent effect. 

Another result shown by the GBM bursts and consistent with the conclusion 
drawn from BATSE is that the ST effect is not responsible for the 
distribution of the data in the plane, i.e. 
it cannot explain why bursts tend to distribute along a correlation. 
This is well visible for BATSE bursts: events with large \epo\ tend to 
concentrate far from the ST, i.e., at higher fluences. 
The trend shown by the BATSE ST (which requires higher limiting fluences when 
\epo\ is very high and very low) cannot explain this behavior. 
The same holds for GBM bursts, for which the ST are even less curved 
and cannot be responsible for the observed correlation which has 
a slope $\sim0.20\pm0.04$, consistent with that defined by BATSE bursts (N08).

\vskip 0.5 cm
\noindent
\emph{\epo\ vs Peak Flux ---}
The distribution of GBM long bursts in the \epop\ plane with respect to BATSE 
bursts is shown in the upper right panel of Fig. \ref{piani correle}. 
Solid curves represent the TT derived for both instruments 
(adapted from Band et al. 2006). 
On average, the GBM instrument is a factor 3 less sensitive than BATSE
in the common energy range. 
As remarked by N08, the sample of BATSE bursts lies far from its TT, 
suggesting that for this instrument the 
demand of performing a reliable spectral analysis is the dominant selection query. 
This is not the case for the GBM: the data points lie very near 
the TT curves, suggesting that if a burst is detected there is a 
good chance to recover its spectral parameters.  
This makes TT and ST competitive selection effects for GBM bursts. 

\subsubsection{Short bursts}

For short bursts (bottom panels in Fig. \ref{piani correle}) the situation is 
quite different in both \epof\ and \epop\ planes than for long events. 

Although GBM short bursts are still only a few (filled circles in Fig. \ref{piani correle}) 
there is a weak indication of a correlation in both planes (left and right bottom panels 
in Fig. \ref{piani correle}), consistent with the trends suggested by BATSE short bursts (open squares). 
However, the overall behavior is different from what happens for long GRBs.

At high \epo\ GBM short events occupy the same region of the BATSE ones, and even 
extend the \epof\ trend  to \epo\ values larger than 1 MeV (i.e. above the BATSE upper threshold), 
revealing that short GRBs with \epo\ larger than 1 MeV exist in the population of GBM events.

Furthermore, contrary to what one might expect, there are no short GBM bursts with 
\epo\ below $\sim 200$ keV. 
At low fluences this can be accounted for by considering the ST derived for the GBM instrument:
Fig. \ref{piani correle} shows that GBM short events lie very near to the ST curve 
(bottom left panel of Fig. \ref{piani correle}), that prevents the estimate of 
\epo $< 200$ keV when $F<4\times10^{-7}$erg cm$^{-2}$.

The higher sensitivity of BATSE instead implies that its ST is located at lower fluences. 
This however does not account for the absence of short GBM burst with \epo$< 200$ keV and  
fluence $> 5\times10^{-7}$ erg cm$^{-2}$: the BATSE sample shows that GRBs with low \epo\ 
but on the right side of the GBM ST do exist. 
Their absence in the present GBM sample seems to imply that they are relatively rare and 
therefore that at a fluence $>4\times 10^{-7}$ erg cm$^{-2}$ most of the events
should have \epo$>$ 200 keV. In other words this would support the presence of a \epo-\F\  
correlation for short GRBs albeit with a large dispersion. 
A larger sample of short GBM bursts is required to confirm this.

In terms of \epo\ this translates into a distribution peaked at higher energies 
compared both to the BATSE one and to the GBM long events (see \S. 4.2).

In the \epop\ plane of short GRBs (right bottom panel in Fig. \ref{piani correle}) 
similar conclusions can be drawn: for both 
instruments the TT curves (solid lines) do not affect the samples. Both samples are 
clearly limited by the selection cut applied on the peak flux (shaded curve).  
From the fact that the peak flux of GBM bursts is significantly above their TT, we infer 
that their selection is dominated by the ST.

\subsection{Outliers of the \ama\ and \yone\ correlations}

Another relevant point is to test whether bursts without measured redshifts 
are consistent with the \ama\ and \yone\ correlations. 
These correlations are defined in the rest frame and require, to add a burst on 
top of them, to have the redshift known.
However, as first proposed by Nakar \& Piran (2005) and then by Band \& Preece (2005), 
knowing \epo\ and the fluence or peak flux it is still possible to test if a burst, 
without measured redshift, is an outlier of the \ama\ and \yone\ correlations. 

In the observational planes it is possible to define a ``region of outliers". 
We start by writing the \ama\ correlation as (the same argument can be repeated for 
the \yone\ correlation): 
\begin{equation}
E_{\rm   peak}=K  E_{\rm  iso}^{\eta}
\end{equation}  
Since \epo$=E_{\rm peak}/(1+z)$ and  $F=E_{\rm   iso}(1+z)/4\pi d_L^2(z)$,  
we can form the ratio,
\begin{equation} 
{(E{^{\rm obs}_{\rm peak}})^{1/\eta}\over F} \, =\,  
K^{1/\eta}\,10^{\pm \sigma/\eta}\, {4\pi d_{L}^{2}(z)\over (1+z)^{(1+\eta)/\eta}} 
\end{equation} 
where $\sigma$ corresponds to the scatter of data points around the rest frame 
correlation that is being tested. 
{\it 
Note that this is not the error on the slope or normalization of 
the correlation, but the scatter measured perpendicular to the best 
fit line of the \ama\ correlation and modeled as a gaussian distribution.  
}

The RHS of the  above relation is a function of $z$, $\eta$ and $\sigma$ only.
The upper limit of the ratio $E_{\rm peak,obs}^{1/\eta}/F$ establishes an allowance region boundary 
on the corresponding plane (upper left--corner shaded regions in Fig. \ref{piani correle}). 
All the bursts that fall below this line in the observational planes can 
have a redshift 
which make them consistent with the \ama\ correlation within its 3$\sigma$ scatter. 
Those falling in the shaded region are outliers at more than 3$\sigma$ for any assigned redshift. 

Although this test has been already applied several times in the recent past, 
some guidelines should be followed: 
\begin{enumerate}
\item
since the rest frame correlations are defined with the bolometric 1 keV--10 MeV 
\eiso\ and \liso, then when testing the region of outliers in the observational planes 
one should use the fluence $F$ and peak flux $P$ defined on the same energy range; 
\item
it is correct to consider the 3$\sigma$ scatter of the rest frame correlations 
and not the uncertainty on the slope ($\eta$) and normalization ($K$) of the correlations. 
This is because the scatter $\sigma$ of the \ama\ and \yone\ correlations  dominates over the 
statistical uncertainty on $K$ and $\eta$ (e.g. G10); 
\item
while the \ama\ and \yone\ correlations were first derived with only a dozen of bursts, 
they have been now updated with nearly 100 GRBs with measured redshifts (e.g. N08, G10): 
the correlation parameters (slope, normalization and scatter) have changed since their discovery, 
so one should adopt the most updated versions of these correlations. 
\end{enumerate}

N08 find that 6\% of BATSE long bursts are outliers of the 
\ama\ correlation, while no outlier is found for the \ep--\liso\ correlation. 
Almost all short \ba\ bursts 
are outliers of the \ep--\eiso\ correlation (defined by long bursts)
but they can be consistent with their very same \ep--\liso\ 
correlation. 
These findings are supported by the consistency of the few short 
bursts with measured redshift with the \yone\ correlation while they are outliers 
at more than 3$\sigma$ of the \ama\ correlation (G09).

For GBM long bursts, Fig. \ref{piani correle} shows that only 8 GRBs
(i.e. 3\%) lie in the region of outliers (in the \epof\ plane), even if the extended energy
range of the GBM allows to explore the region of large \epo\ and intermediate
fluences, where outliers, if they exist, could be found.

Instead, for short GBM bursts, we can see that most of them
are outliers at more than 3$\sigma$ of the \ama\ correlation.
On the contrary, there are no short bursts in the region of outliers in 
the \epop\ plane, i.e. they are all consistent with the \yone\ correlation defined by long GRBs. 
The hypothesis that short and long bursts follow the same \ep--\liso\ 
correlation is also supported by the few short events with known redshift (G09).

\section{Spectral parameter distributions}

In this section we compare the distributions of the spectral parameters 
(low energy spectral index $\alpha$ and peak energy \epo) for 
long (\S 4.1) and short (\S 4.2) GRBs detected by BATSE and by the GBM.
In addition, we show the Log$N$--Log$P$ distribution for short bursts.

\subsection{Long Bursts}

\begin{figure} 
\vskip -0.3 cm
\includegraphics[scale=0.55]{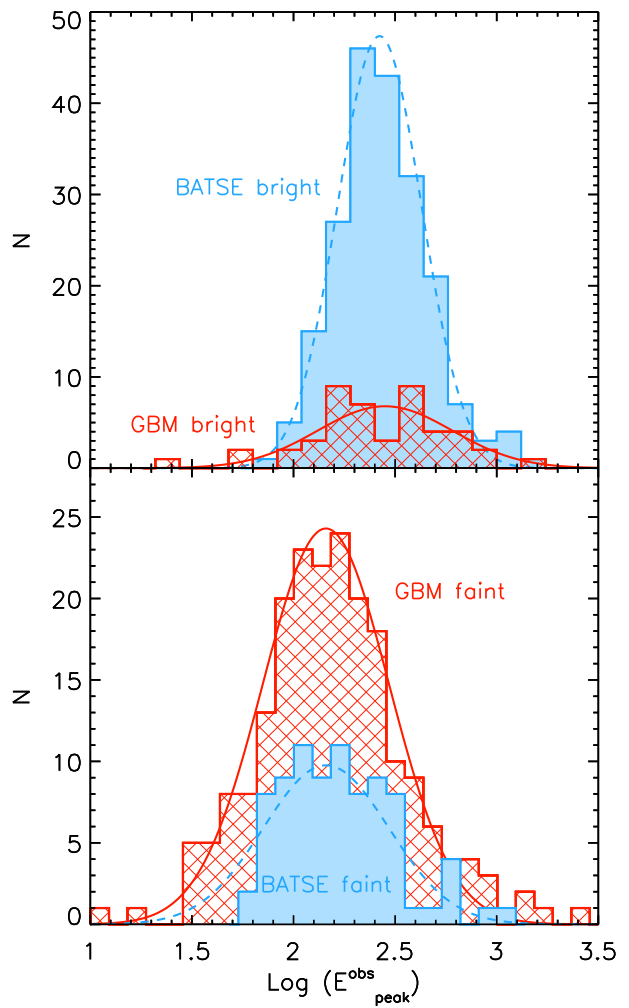}
\caption{
{\it Long GRBs.}
\epo\ distribution for BATSE bursts (blue solid filled histograms) and GBM bursts 
(red hatched histograms). 
For both instruments, we plot separately the \epo\ distributions for 
the bright sample (upper panel) and the faint sample (bottom panel).
}
\label{ep_long} 
\end{figure} 

\begin{figure} 
\vskip -0.2 cm
\includegraphics[scale=0.55]{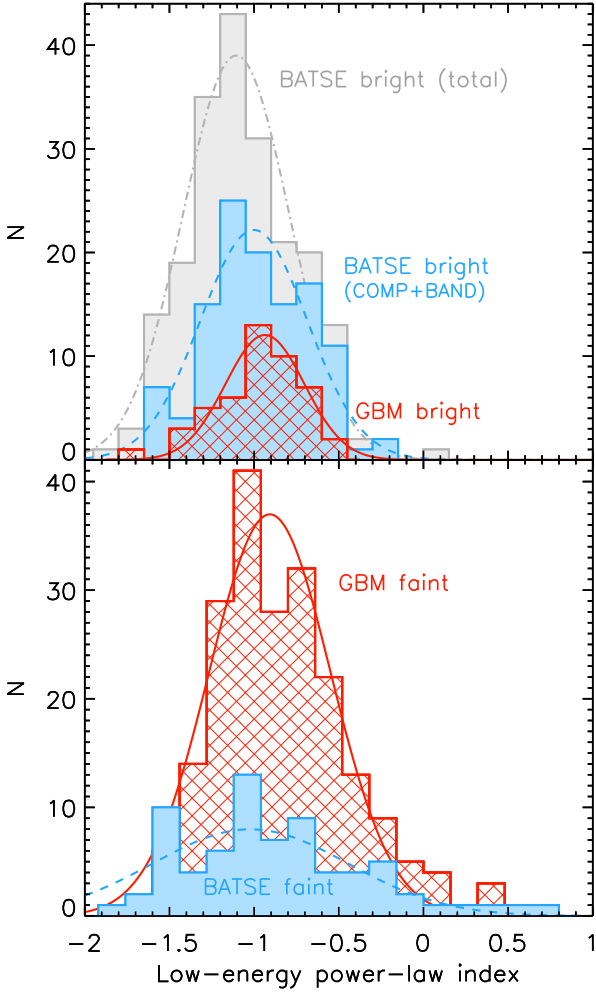}
\caption{
{\it Long GRBs.}
Low--energy power law index $\alpha$ distributions for BATSE bursts (blue solid filled histograms) 
and GBM bursts (red hatched histogram). 
For both instruments, we plot separately the 
distributions for the bright sample (upper panel) and the faint sample (bottom panel). 
For the bright \ba\ sample of K06 (upper panel) we also show separately the $\alpha$ 
distribution of bursts fitted with all models (light gray shaded histogram) and that 
of bursts fitted with only the COMP or Band model (dark blue histogram).} 
\label{a_long} 
\end{figure} 

Fig. \ref{ep_long} (upper panel) shows the distributions of \epo\ of bright BATSE 
and bright GBM bursts. 
The two distributions are quite similar: 
the central value of the gaussian fit (solid line) is \epo$\sim$260 keV and \epo$\sim$280 keV, 
respectively, with GBM bursts having a larger distribution (standard deviation 
$\sigma$=0.33 to be compared with $\sigma$=0.21 for \ba\ bursts) and extending 
both at lower and higher \epo with respect to that of \ba. Using the Kolmogorov--Smirnov (KS) 
test, we find that the probability that the two distributions are drawn from the same 
parent population is 0.164. 

For faint bursts (bottom panel of Fig. \ref{ep_long}) the results are similar. 
For both instruments the gaussian fit to the \epo\ distribution is centered around 
140 keV and has a standard deviation $\sigma$=0.31 (the KS test probability is 0.18). 
Also in this case the GBM distribution is larger. 
In particular, from the 
comparison of the two histograms it is appears that GBM data allows to recover very 
low \epo, thanks to the good GBM/NaI sensitivity down to 8 keV (i.e. extending by a 
factor of 3 the low--energy bound of \ba). 
For BATSE bursts there is a quite sharp cutoff at $\sim$ 50 keV. 
This is in agreement with the simulations performed by G08 
(see Fig. \ref{piani correle}) showing that it is very difficult for \ba\ to 
recover \epo\ $<$ 50 keV. 
Large fluences, unusual for such values of \epo, would be required. 
All the results of the gaussian fits and KS probabilities are summarized in Tab. \ref{numeri}.

We stress that the \epo\ properties of faint and bright bursts are very different 
due to the correlation between \epo\ and $F$: the faint sample is characterized by a 
central value of \epo\ which is almost a factor of 2 lower than that of  the bright sample. 
The KS probability of the distributions of \epo\ within the GBM sample  
between faint and bright bursts is 5.7$\times10^{-5}$.

Fig. \ref{a_long} shows the $\alpha$ distribution for bright (upper panel) and faint  
burst (bottom panel). 
In both cases, GBM bursts tend to have a harder low--energy 
power law index and a somewhat tighter distribution (central values and standard 
deviations of the gaussian fits are reported in Tab. \ref{numeri}). 
However the KS test shows that the $\alpha$ distributions of
faint GBM and faint \ba\ bursts are similar (KS probability=0.12). 
Moreover, GBM bursts do not show a significant relation between $\alpha$ and the fluence 
because faint and bright GBM bursts have similar distributions peaked respectively 
at --0.91 and --0.94 and with a KS probability of 0.03.

The $\alpha$ distribution of bright \ba\ bursts (grey histogram), instead, shows significant difference 
with respect to both the bright GBM sample (KS probability=6$\times10^{-3}$) and the 
faint \ba\ sample (KS probability=7$\times10^{-3}$). 
Bright \ba\ bursts tend to have 
softer $\alpha$ with respect to all the other samples. 
We investigated the possible 
origin of this difference considering that the K06 sample contains bursts whose spectra 
are fitted with the SBPL, COMP or Band model. 
As noted by K06 themselves the spectral 
parameters \epo\ and $\alpha$ do show different typical values and width of 
their distributions depending on the fitting spectral model. 
Considering that GBM 
bursts are adequately fitted by either the COMP or the Band model, we  excluded 
from the \epo\ distribution of \ba\ bursts the bursts fitted with a SBPL. 
The resulting histogram (dark blue shaded in Fig. \ref{a_long}) is now fully 
consistent with the distribution of $\alpha$ of GBM bursts 
(the KS probability now becomes 0.4).

\subsection{Short Bursts}

\begin{figure} 
\vskip -0.5 cm
\hskip -1.12 cm
\includegraphics[scale=0.555]{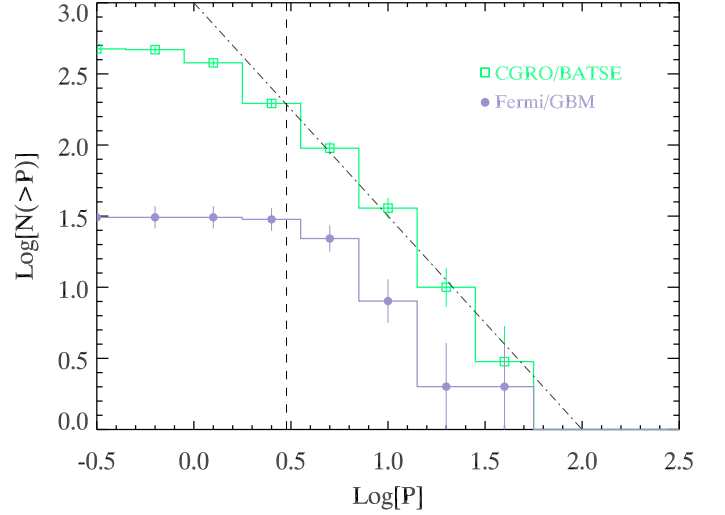}
\vskip -0.2 cm
\caption{
{\it Short GRBs.} 
Log$N$--Log$P$ for short \ba\ bursts (empty squares) and short GBM bursts (filled circles). 
In both cases the peak flux $P$ (in photons cm$^{-2}$ s$^{-1}$) is integrated between 50 keV 
and 300 keV on a time scale of 64 ms. 
GBM data are taken form N10, while \ba\ data are 
from the online catalog and include all the \ba\ bursts for which the peak flux has been estimated. 
The vertical dashed line is the flux limit of the selection of \ba\ bursts analyzed in G09 and for reference a power law 
with slope --3/2 is shown (dot--dashed line). 
}
\label{pf_short} 
\end{figure} 

The largest sample of \ba\ bursts for which the spectral analysis has been performed 
was selected on the basis of a peak flux criterion (G09). 
A meaningful comparison 
between GBM and \ba\ short bursts requires a sample of short GBM bursts selected on 
the basis of the very same criterion. 
Before investigating the spectral parameter 
distributions, we compare the Log$N$--Log$P$ for both instruments, where $P$ in this case is the peak flux in photons cm$^{-2}$ s$^{-1}$. 
Since for the GBM sample N10 estimate the peak flux on the 64 ms timescale, also 
for \ba\ bursts we select (from the online 
catalog\footnote{http://heasarc.gsfc.nasa.gov/W3Browse/cgro/batsegrb.html}) 
all bursts for which the $P$ on 64 ms has been estimated. 
For \ba\ bursts the peak flux is integrated in the 50 keV -- 300 keV energy range. 
Therefore, we estimate for all short GBM bursts in N10 the photon peak flux between 50 keV and 300 keV. 
Fig. \ref{pf_short} shows our results. 

As discussed in \S 3 the lack of bursts with low peak flux in the GBM sample is 
due to the ST threshold shown in Fig. \ref{piani correle}. 
This instrumental threshold, 
indeed, dominates over the TT threshold and determines that short GRBs for which the 
spectrum can be analyzed and the spectral parameters properly constrained should 
have a large number of photons. At high peak fluxes instead, GBM has detected 
more short GRBs than \ba\ due to the \epop\ correlation, which associates high 
peak fluxes to high peak energies, the latter better constrained with the larger 
energy range of the GBM instrument (BGO) than with \ba. 
This explains the different shapes of the two Log$N$--Log$P$. 

\begin{figure} 
\includegraphics[scale=0.55]{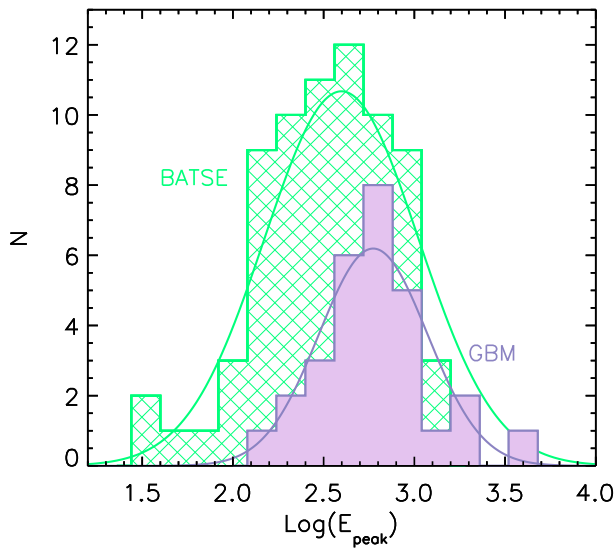}
\vskip -6. cm
\caption{
{\it Short GRBs.} 
\epo\ distribution for BATSE bursts (green hatched histogram, from G09), 
and GBM bursts (purple filled histogram, from N10). 
} 
\label{ep_short} 
\end{figure}

To compare the spectral parameters, we consider short GRBs (from G09 for \ba\ and N10 for GBM)  
with known $\alpha$ and \epo\ and with $P$(50--300 keV)$>$3 ph cm$^{-2}$ s$^{-1}$. 
The \epo\ distributions are shown in Fig. \ref{ep_short} and are quite different. 
The lack of low \epo\ in the GBM sample (which corresponds to the lack of low peak 
fluxes in Fig. \ref{ep_short}) can be explained by considering the shape and position of 
the ST for short bursts (left bottom panel in Fig. \ref{piani correle}) and 
the existence of a correlation between $P$ and \epo\ (see discussion above and in \S 2). 
Moreover, contrary to long bursts, the \epo\ distribution of GBM short events extends 
to higher energies, suggesting that such large values of \epo\ can be found in  short 
bursts and that they were not present in the \ba\ catalog due to its limited energy 
range (up to only  $\sim$ 1 MeV).

The $\alpha$ distributions of \ba\ and GBM bursts is considerably different. 
The GBM confirms that short events are harder than long ones in terms of low--energy 
spectral index (KS probability=3$\times10^{-8}$). However, the distribution is 
peaked around $\alpha = -0.59$ and it is very narrow ($\sigma$=0.15). 
We tentatively interpret this as due to the large energy range 
of GBM which extends down to 8 keV, 
but this point deserves further study.
The extension down to low energies of GBM allows in principle to determine $\alpha$
more accurately. 
Instead, the limited energy range of \ba\ resulted in less accurate 
estimates of $\alpha$ and thus a more dispersed distribution of its values. This effect is slightly present 
also in long bursts, both bright and faint (see Tab. \ref{numeri}).

\begin{figure} 
\includegraphics[scale=0.55]{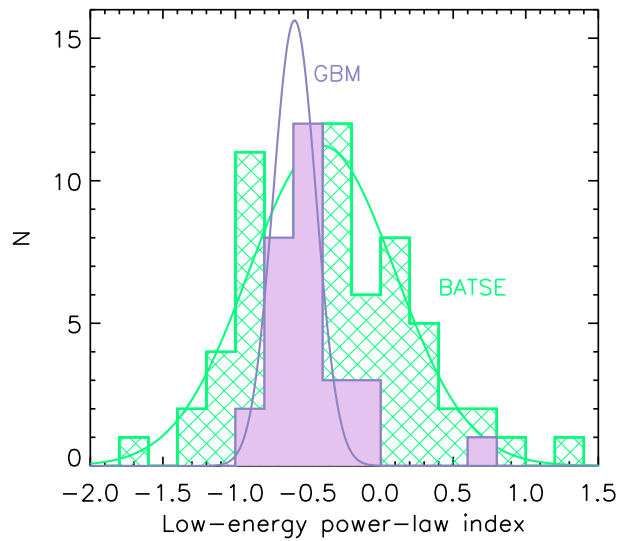}
\vskip -6 cm
\caption{
{\it Short GRBs.} 
$\alpha$ distribution for BATSE bursts (green hatched histogram, from G09), 
and GBM bursts (purple filled histogram, from N10).
} 
\label{alpha_short} 
\end{figure} 

\begin{table*}
\begin{center}
\begin{tabular}{c|ccc|ccc|ccc}
\hline
\hline
\multicolumn{1}{c|}{} & \multicolumn{6}{c|}{LONG} & \multicolumn{3}{c}{SHORT}\\
\multicolumn{1}{c|}{Param.} & \multicolumn{3}{c|}{Bright} & \multicolumn{3}{c|}{Faint} & \multicolumn{3}{c}{}\\

      & BATSE & Fermi & KS &   BATSE & Fermi & KS &  BATSE & Fermi & KS   \\
\hline
E$_{\rm peak}$  & 2.42   & 2.45  &  0.164   & 2.16 & 2.16 & 0.18   & 2.60 & 2.77 & 3.6$\times 10^{-3}$ \\
                          &(0.21)  &(0.33) &              &(0.31)&(0.31)&          &(0.42)&(0.30)&                                 \\
\hline
Alpha                 &--1.00 &--0.94 &   0.401   &--1.02&--0.91& 0.12 &--0.40&--0.59 & 0.018 \\  
                          &(0.31) &(0.23) &               &(0.57)&(0.36)&         &(0.50)&(0.15) & \\
\hline
\hline
\end{tabular}
\caption{
Central values and standard deviations (in brackets) of the distributions 
of $\alpha$ and \epo\ for long and short bursts. The table also lists the KS 
probability resulting from the comparison between BATSE and GBM distributions 
of $\alpha$ and \epo. For long bright bursts the comparison has been performed by considering (for homogeneity) 
bursts best modeled by a COMP or a Band model (i.e. by excluding from K06 those bursts modeled with a SBPL).}
\label{numeri}
\end{center}
\end{table*}

\section{Discussion and Conclusions}
\begin{figure*}
\vskip -1 true cm
\includegraphics[scale=0.6,angle=0]{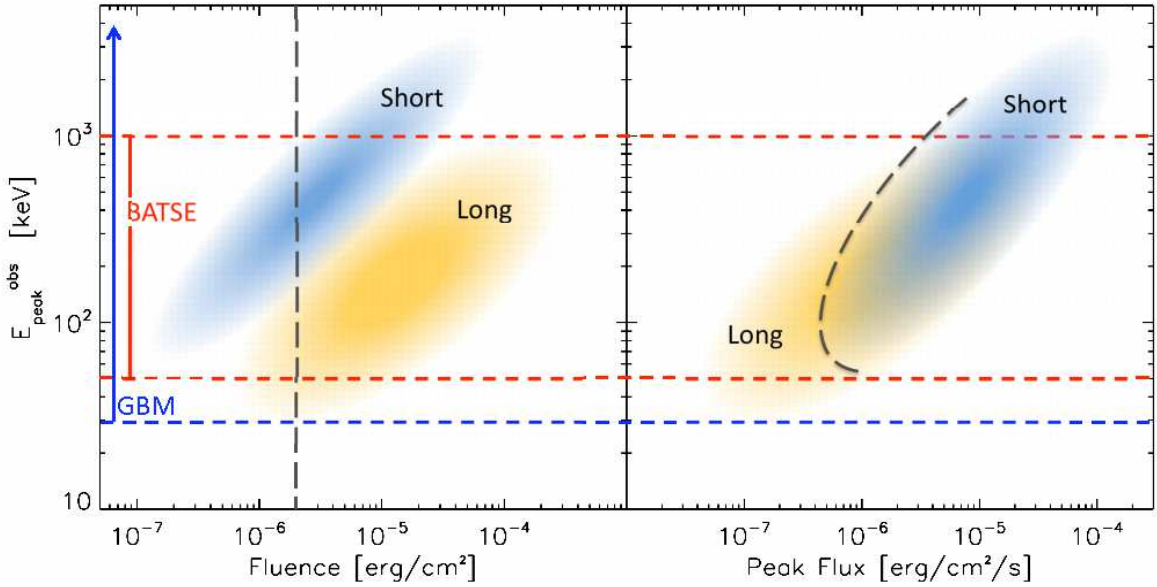}
\vskip -2 true cm
\caption{
Schematic view of the distribution of long and short GRBs in the \epof\ and \epop\ planes. 
The horizontal dashed line at $\sim$ 30 keV represents 
the lower limit for the GBM instrument: the simulations performed in this work 
show that \epo\ can be hardly determined below this value. 
For the BATSE instrument this limit corresponds to $\sim$ 50 keV. 
The upper limit for BATSE is at $\sim$ 1 MeV, while for the GBM there is no 
upper limit in this plane. 
The vertical dashed line (left panel) shows an example of fluence selection, 
while the dashed curve (right panel) refers to the photon flux selection 
criterion adopted by G09.}

\label{figurella} 
\end{figure*} 

In this work we investigated the presence of the \epo--Fluence and \epo--Peak Flux 
correlations in GBM bursts (both long and short) detected by the GBM instrument up to March 2010. 
Similarly to what has been done for long and short GRBs detected by 
BATSE (N08; G09) we examined the distribution of GBM bursts in the 
\epo--Fluence and \epo--Peak Flux planes in order to study instrumental selection effects 
and test their consistency with the 
rest frame correlations (i.e. \ama\ and \yone, respectively) defined by 
GRBs with measured redshifts. 
To this aim, we have estimated, for the GBM instrument, the 
spectral threshold (ST) and the trigger threshold (TT), in order to quantify the 
selection effects acting on the considered samples and their role on the found correlations. 

Our main results are:
\begin{itemize}

\item 
{\it Long GRBs} detected by GBM follow the same \epof\ and \epop\ 
correlations defined by BATSE GRBs (Fig. \ref{piani correle}). 
We computed the instrumental selection effects of GBM -- as already done for BATSE 
(G08; N08): the trigger threshold and the spectral threshold are not 
responsible for the correlations defined by long GRBs in both planes 
(see Fig. \ref{piani correle}).
The GBM spectral extension down to 8 keV with respect to the limit of 
30 keV of BATSE,  allows to extend the correlations to lower peak energies/fluences.
Instead, despite of the higher energy threshold of 
GBM (40 MeV) no long GRB with \epo\ larger than a few MeV is detected. 
This can be due to a real absence of bursts with such high \epo\ or to the fact 
that they have large fluences, thus being too rare to be detected 
during less then 2 years of the GBM observations.

We conclude that long GRBs detected by GBM confirm what 
found with BATSE bursts, in particular that they follow a correlation both 
in the \epof\ and in the \epop\ plane. 
Moreover, the fraction of bursts detected by GBM which are outliers at more 
than 3$\sigma$ with respect to the \ama\ correlation is $\sim$3\%, to be compared with the 6\% 
of outliers found (N08) in the  BATSE sample. 
Instead, there are no outliers (at more than 3$\sigma$) of the \yone\ correlation 
among GBM long GRBs.

\item {\it Short GRBs} detected by GBM populate a different region in 
the \epof\ plane with respect to long events, the former having larger peak 
energies and lower fluences compared to the latter. 
This is consistent with what found by BATSE and confirms that short GRBs do not follow 
the ``Amati" correlation but they obey the ``Yonetoku" correlation 
defined by long events.
\end{itemize}

The GBM population of long and short bursts with spectral information is large enough 
to allow a statistical comparison with the BATSE results. 
For long bursts, we considered 
the fluence distribution of \ba\ bursts and we compare it to those derived by N10 for GBM bursts. 
We also compared the spectral properties for selected samples of GBM and \ba\ bursts 
with well defined \epo\ derived from the spectral analysis. 
Two different sample of \ba\ bursts are available in literature, 
based on complementary fluence selection criteria. 
We call them faint and bright \ba\ samples. 
We then selected from the catalog of N10 
two subsamples of GBM bursts based on the same fluence criteria applied to the \ba\
samples (i.e. $10^{-6}$ erg/cm$^2 < F < 2\times10^{-5}$ erg/cm$^2$ for the faint 
GBM sample and $F>  2\times10^{-5}$ erg/cm$^2$ for the bright GBM sample).

The \epo\ distribution derived from the two instruments are quite similar (Fig. \ref{ep_long} and Tab. \ref{numeri}). 
Despite its larger energy range, the GBM extends the \epo\ distribution 
of long bursts only at low energies with respect to BATSE. 
The $\alpha$ distribution, 
instead, reveals some difference for the sample of bright bursts: \ba\ bursts have on average a softer 
low--energy photon index ($\langle\alpha_{\rm GBM}\rangle=-0.9$ and 
$\langle\alpha_{\rm BATSE}\rangle=-1.11$, KS probability $=6\times10^{-3}$). However,
this difference is almost totally due to the presence (in the K06 sample of bright BATSE bursts) of GRBs modeled 
by a smoothly broken power--law (SBPL) function. As noted by K06, this model gives a low--energy spectral index systematically softer
with respect to COMP and Band models. By excluding these events, the $\alpha$ distribution of bright BATSE bursts
is centered around $\langle\alpha_{\rm BATSE}\rangle=-1.00$ and the KS probability with the GBM is 0.4.

Also for GBM short bursts we can draw some conclusions about their spectral properties. 
Their \epo\ distribution is shifted towards higher energies compared both 
to long bursts from the same instrument and to short bursts seen by BATSE. 
The lack of low--energy \epo\ (below $\sim 200$ keV) can be accounted 
for by the spectral threshold we derived for the GBM instrument 
(see Fig. \ref{piani correle}). This hypothesis is supported by the fact that among the population of short GBM bursts 
there are 44 events fitted with a curved model (i.e. with \epo\ determined) but there exists a large fraction of short
bursts (34) whose spectrum is fitted with a single power law. 
On the other hand, the larger energy 
coverage allows the detection of \epo\ up to $\sim$ 4 MeV. 
GBM data confirm that short bursts have on average a harder $\alpha$ 
compared to long bursts ($\langle\alpha_{\rm GBM,short}\rangle\sim -0.59$),
as already found in the BATSE sample by G09.

The comparison of short and a representative sample of long BATSE GRBs 
(selected with a similar peak flux threshold) led G09 to conclude 
that their main spectral diversity is due to a 
harder low--energy spectral index in short bursts while their 
\epo\ of BATSE is similarly distributed. 
GBM bursts provide the opportunity of re--examining this result 
for the population of short and long GRBs detected by the GBM and 
also compare their spectral properties with those of the  BATSE ones. 
We find that:

\begin{itemize}
\item 
\epo\ of short GBM bursts is larger and \al\ smaller that those of long one, indicating that 
short events are harder, both in terms of their peak energy and low--energy spectral index. 

\item 
A comparison between GBM and BATSE short bursts reveals that they 
have similar \al\ while the \epo\ of short GBM bursts is larger 
than that of short BATSE events (see Fig. \ref{piani correle}, bottom left panel). 
This information is allowed by the higher energies which can be detected by the GBM.
Moreover, the different \epo\ distribution of BATSE and GBM short 
bursts is affected by the lower sensitivity of the GBM instrument, which 
misses short bursts at low fluences (and therefore low \epo).

\item 
GBM and BATSE long bursts have a similar \epo\  while GBM events tend to 
have a harder low--energy spectral index (Fig. \ref{a_long}).

\end{itemize}

Fig. \ref{figurella} shows a schematic representation of the current 
information about the distribution of short and long bursts in the \epof\ and 
\epop\ planes. 
With respect to BATSE, the GBM reveals that long bursts 
extend to lower \epo, consistently with what previously found with other 
instruments (mainly {\it Hete--II} and {\it Swift}). 

Despite of the high--energy sensitivity, also the \epo\ distribution of GBM long events
extends only up to $\sim$ 1 MeV. The situation is different for short GRBs
whose \epo\ reach up to $\sim$ 4 MeV in the present sample. 
These high \epo\ were  
not detectable by BATSE, whose sensitivity drops at $\sim$ 1 MeV (upper 
horizontal dashed line in Fig. \ref{figurella}). Therefore, 
the GBM shows that short GRBs have larger \epo\ with respect to long ones, 
contrary to what found with BATSE (G09). 

When comparing the \epo\ distribution of short and long bursts, different 
conclusions can be drawn, according to the selection criterion of the samples. 
The left panel in Fig. \ref{figurella} shows that 
a given cut in fluence (represented 
by the vertical dashed line) would result in different \epo\ distributions between short 
and long bursts, resulting from their different location in the \epof\ plane. 
The right panel in Fig. \ref{figurella} illustrates, instead, what happens for 
a selection in {\it photon flux}. 
This translates into a curve in {\it energy flux}: 
the dashed curve represents the cut 
applied by G09 to select both short and long bursts, corresponding to 
a photon flux larger than 3 ph cm$^{-2}$s$^{-1}$. 
This criterion applied to BATSE bursts produces similar \epo\ distributions 
of long and short events, as indeed found by G09. 
The very same criterion applied to GBM bursts results, instead, in different 
distributions, since short bursts can have very 
high \epo\ values, not detected in the sample of long bursts.

\begin{acknowledgements} 
We thank Y. Kaneko for private communications on the BATSE bright sample spectral results. 
This research has made use of data obtained through the high--energy Astrophysics Science Archive 
Research Center Online Service, provided by the NASA/Goddard Space Flight Center. 
This work has been partly supported by ASI grant I/088/06/0. LN thanks the 
Osservatorio Astronomico di Brera for the kind hospitality for the completion of this work. 
\end{acknowledgements}

\bibliographystyle{aa}

\end{document}